\documentclass[a4paper,fleqn,usenatbib]{mnras}

\usepackage[T1]{fontenc}
\usepackage{ae,aecompl}

\usepackage{graphicx}	
\usepackage{amsmath}	
\usepackage{amssymb}	

\usepackage[usenames,dvipsnames]{color}

\DeclareMathAlphabet{\mathpzc}{OT1}{pzc}{m}{it}
 
\def\beq{\begin{equation}}
\def\eeq{\end{equation}}
\def\bea{\begin{eqnarray}}
\def\eea{\end{eqnarray}}
\def\bwt{\begin{widetext}}
\def\ewt{\end{widetext}}

\def\nn{\nonumber\\}

\title[Which is a better cosmological probe: Number counts or cosmic magnification?]{Which is a better cosmological probe: Number counts or cosmic magnification?}

\author[D.~Duniya \& M.~Kumwenda]{
Didam G. A. Duniya$^{1}$\thanks{E-mail: duniyaa@biust.ac.bw}
and
Mazuba Kumwenda$^{1,2}$
\\
$^{1}$Department of Physics \& Astronomy, Botswana International University of Science and Technology, Palapye, Botswana\\
$^{2}$Department of Physics, The Copperbelt University, Kitwe, Zambia\\
}

\date{Accepted 2023 April 21. Received 2023 April 11; in original form 2023 February 17}

\pubyear{2023}

\begin{document}
\label{firstpage}
\pagerange{\pageref{firstpage}--\pageref{lastpage}}
\maketitle

\begin{abstract}\noindent
The next generation of cosmological surveys will have unprecedented measurement precision, hence they hold the power to put theoretical ideas to the most stringent tests yet. However, in order to realise the full potential of these measurements, we need to ensure that we apply the most effective analytical tools. We need to identify which cosmological observables are the best cosmological probes. Two commonly used cosmological observables are galaxy redshift number counts and cosmic magnification. Both of these observables have been investigated extensively in cosmological analyses, but only separately. In the light of interacting dark energy (IDE) emerging as a plausible means of alleviating current cosmological tensions, we investigate both observables on large scales in a universe with IDE, using the angular power spectrum: taking into account all known terms, including relativistic corrections, in the observed overdensity. Our results suggest that (given multi-tracer analysis) measuring relativistic effects with cosmic magnification will be relatively better than with galaxy redshift number counts, at all redshifts $z$. Conversely, without relativistic effects, galaxy redshift number counts will be relatively better in probing the imprint of IDE, at all $z$. At low $z$ (up to around $z \,{=}\, 0.1$), relativistic effects enable cosmic magnification to be a relatively better probe of the IDE imprint; while at higher $z$ (up to $z \,{<}\, 3$), galaxy redshift number counts become the better probe of IDE imprint. However, at $z \,{=}\, 3$ and higher, our results suggest that either of the observables will suffice.
\end{abstract}




\section{Introduction}\label{sec:intro}
As we enter the era of precision cosmology, where surveys will have unprecedented measurement ability, observational data will hold unmatched power to test theoretical ideas. This presents a strong opportunity for innovative work. It will be possible to test, and to distinguish more precisely than before, which of the observables are better or even the best cosmological probes. For example, we will be able to answer the crucial question of whether galaxy redshift number counts (i.e. the number counts of galaxy redshift surveys) or cosmic magnification, is a better cosmological probe. It is important to answer this question since, in order to realise the full potential of the forthcoming observational data, we need to employ the most effective analytical tools. However, in order to properly answer the given question, a rigorous quantitative analysis is required. Such analysis needs to take into account several factors: Firstly, the survey specifications such as survey area, duration of survey, kind of survey (spectroscopic or photometric), and so on. Secondly, the signal considered, e.g. the matter density amplitude, the baryon acoustic oscillations \citep[see e.g.][]{BOSS:2012dmf}, the primordial non-Gaussianity \citep[see e.g.][]{Hu:2001url, Celoria:2018euj, Byrnes:2010em, Maartens:2012rh}, and so on. Thirdly, the statistic used for the analysis, e.g. the two-point correlation function, the linear power spectrum, the angular power spectrum, the three-point correlation, and so on. Fourthly, error statistics, including signal-to-noise ratios and the cosmic variance \citep[see e.g.][]{Maartens:2012rh, Duniya:2019mpr}.

The galaxy redshift number counts measure the distribution of the galaxies in the sky. Astronomers study the properties of this distribution and the implications for the content and the evolution of the Universe: the galaxy distribution maps the underlying matter---up to a ``bias'' factor~\citep{Baldauf:2011bh, Bartolo:2010ec, Jeong:2011as, Lopez-Honorez:2011emg, Duniya:2016ibg, Desjacques:2016bnm}---which is a key pointer to the details of the past, the present, and the future of the Universe. On large (linear) scales, the matter distribution provides a clear, relatively easy window back to the cosmic primordial conditions. On small (non-linear) scales, it is relatively difficult to extract information about the initial fluctuations. Essentially, galaxy redshift number counts are a diagnostic of the growth of structure; hence, are able to probe the nature of dark energy (DE) \citep[see e.g.][]{Amendola:2010bkk, Chevallier:2000qy, Linder:2002et, Duniya:2013eta, Duniya:2015nva, Duniya:2019mpr, Duniya:2015dpa}. The analysis of the galaxy distribution may be done via the observed number-count overdensity \citep{Yoo:2009au, Yoo:2010ni, Yoo:2014kpa, Challinor:2011bk, Bonvin:2011bg, Jeong:2011as, Lopez-Honorez:2011emg, Maartens:2012rh, Bonvin:2014owa, Bonvin:2015kuc, Gaztanaga:2015jrs, Duniya:2016ibg, Durrer:2016jzq} of galaxy redshift surveys. (Henceforth, we use ``number-count overdensity'' to mean number overdensity of intrinsically bright sources, such as those observed in galaxy redshift surveys.) 

The angular power spectrum of the observed number-count overdensity is widely used as a cosmological probe, and has been used to hunt down the imprint of DE and relativistic effects \citep[see e.g.][]{Yoo:2009au, Yoo:2010ni, Yoo:2014kpa, Bartolo:2010ec, Challinor:2011bk, Bonvin:2011bg, Jeong:2011as, Baldauf:2011bh, Lopez-Honorez:2011emg, Maartens:2012rh, Raccanelli:2013gja, Duniya:2013eta, Duniya:2015nva, Bonvin:2014owa, Alonso:2015uua, Alonso:2015sfa, Fonseca:2015laa, Bonvin:2015kuc, Duniya:2016ibg, Duniya:2015dpa, Durrer:2016jzq, Gaztanaga:2015jrs, Witzemann:2018cdx, Duniya:2019mpr} in the large-scale structure. However, this approach appears to be unsatisfactory, given that the imprint of the cosmological phenomena seem to be insignificant or degenerate in the angular power spectrum \citep[see e.g.][]{Maartens:2012rh, Alonso:2015uua, Duniya:2019mpr}. An alternative, is the cosmic magnification \citep[see e.g.][]{Blain:2001yf, Raccanelli:2013gja, Raccanelli:2016avd, Menard:2002vz, Menard:2002da, Bonvin:2008ni, Ziour:2008awn, Schmidt:2009rh, Schmidt:2010ex, Jeong:2011as, Camera:2013fva, Liu:2013yna, Hildebrandt:2015kcb, Duniya:2016ibg, Duniya:2016gcf, Bacon:2014uja, Bonvin:2016dze, Chen:2018hil, Andrianomena:2018aad, Coates:2020jzw, LoVerde:2006cj, Ballardini:2018cho}. In contrast to galaxy redshift number counts, cosmic magnification is a diagnostic of cosmic distances and sizes; consequently, a diagnostic of the geometry of the large scale structure. Thus, it is also able to probe the nature of DE. 

Cosmic magnification (or demagnification) manifests in the large scale structure by the enhancement (or depletion) of the observed flux of sources: sources are inherently magnified or demagnified in an inhomogeneous universe. There are several complementary phenomena that result in cosmic magnification; these include, weak lensing \citep{Blain:2001yf, Raccanelli:2013gja, Menard:2002vz, Menard:2002da, Bonvin:2008ni, Ziour:2008awn, Schmidt:2009rh, Schmidt:2010ex, Liu:2013yna, Hildebrandt:2015kcb, Camera:2013fva}, time delay \citep{Raccanelli:2013gja}, Doppler effect \citep{Bonvin:2008ni, Bacon:2014uja, Raccanelli:2016avd, Bonvin:2016dze, Chen:2018hil, Andrianomena:2018aad, Coates:2020jzw}, integrated Sachs-Wolfe (ISW) effect \citep{LoVerde:2006cj, Ballardini:2018cho}, and gravitational potential-well effect. (Similar effects manifest in galaxy redshift number counts.) All these phenomena will need to be taken into account in the observed magnification overdensity, in order to realise the true potential of the cosmic magnification as a cosmological probe. The cosmic magnification provides a mostly independent analytical framework. It will be crucial to understanding cosmic distances, and hence the geometry of the large scale structure. Moreover, it will be key in interpreting the data from future surveys that depend on the apparent magnitude or the angular size of the sources, e.g. neutral hydrogen surveys of the SKA \citep[see e.g.][]{Maartens:2015mra} and the baryon acoustic oscillation measurements of BOSS \citep[see e.g.][]{BOSS:2012dmf}. In fact, the cosmic magnification may be used to put constraints on cosmological parameters \citep[see e.g.][]{Menard:2002vz, Menard:2002da}.

In this paper, we seek insights into whether galaxy redshift number counts or cosmic magnification, is the better as a cosmological probe: by investigating relativistic effects and the imprint of DE in the angular power spectrum, on large scales. For both galaxy redshift number counts and cosmic magnification, we have taken care to include all the known signals including the matter density amplitude, weak lensing, redshift space distortions, Doppler distortion, time delay, and ISW distortion, in the angular power spectra. Our choice of the angular power spectrum as preferred statistic lies on two key advantages \citep{Bonvin:2014owa}: (1) Its measurement does not require information of the underlying cosmology, which thus allows observational data to be confronted with theoretical predictions without ambiguities; (2) It is better fit to measure correlations at large angular separations, since the expansion in spherical harmonics inherently includes wide-angle effects. Consequently, the angular power spectrum is well suited for probing both relativistic effects and the imprint of DE: both become substantial on large scales. Moreover, in the light of an observational evidence \citep{Ferreira:2014jhn} for the existence of interacting DE (IDE), and the promising possibility of it providing a plausible solution \citep[see e.g.][]{Pandey:2019plg, DiValentino:2019ffd, Zhai:2023yny} to some of the current cosmological tensions \citep[see e.g.][]{Verde:2019ivm, DiValentino:2020vvd, DiValentino:2020zio, DiValentino:2022fjm, Abdalla:2022yfr}, we consider a universe with an interacting dark sector: where (cold) dark matter and DE exchange energy and momentum in a (non-gravitational) reciprocal manner. 

The main goal of this work is to provide a qualitative analysis of the relative investigative power of galaxy redshift number counts and cosmic magnification as cosmological probes---in the light of future surveys of e.g. the SKA, BOSS, DES~\citep{DES:2005dhi}, PanStars \citep{Chambers:2016jzn}, and Euclid~\citep{EUCLID:2011zbd, Euclid:2019clj} (among others). Although, as previously pointed out, in order to adequately answer the question posed by the title of this paper a comprehensive quantitative analysis will be required (which is outside the scope of the current work), we retain the given title for the sake of arousing discussion in the community and drawing attention to the need for a definitive answer. We start by describing the observed galaxy number-count overdensity in Sec.~\ref{sec:Delta_n}. We discuss the observed magnification overdensity in Sec.~\ref{sec:Delta_M}; outline the dark sector parameters in Sec.~\ref{sec:Dark}; investigate galaxy redshift number counts and cosmic magnification---using the angular power spectrum---in Sec.~\ref{sec:Probes}, and conclude in Sec.~\ref{sec:Concl}.


\section{The Number-count Overdensity}\label{sec:Delta_n}

The galaxy number counts is an important tool in cosmology. It allows comologists to probe the growth of structure in the Universe; consequently, probe the nature of DE, on ultra-large scales (i.e. near and beyond the Hubble horizon). By using the observed number distribution of galaxies in a given direction ${-}{\bf n}$ at redshift $z$, cosmologists are able to construct the galaxy number-count density perturbation (or overdensity), which describes the fluctuation in the density of the number distribution. As stated in Sec.~\ref{sec:intro}, in this work we reserve the term ``number-count overdensity'' to denote the number overdensity of intrinsically bright sources: as measured in galaxy redshift surveys.

Thus, the observed (relativistic) number-count overdensity \citep{Yoo:2009au, Yoo:2010ni, Yoo:2014kpa, Challinor:2011bk, Bonvin:2011bg, Jeong:2011as, Lopez-Honorez:2011emg, Alonso:2015uua, Bonvin:2014owa, Bonvin:2015kuc, Gaztanaga:2015jrs, Duniya:2016ibg, Durrer:2016jzq} of galaxy redshift surveys, is given by 
\bea\label{Delta_n}
\Delta^{\rm obs}_n({\bf n},z_S) \;=\; \Delta^{\rm std}_n({\bf n},z_S) + \Delta^{\rm rels}_n({\bf n},z_S),
\eea
where here we define the ``standard'' term as
\begin{align}\label{stdDelta_n}
\Delta^{\rm std}_n({\bf n},z_S) \equiv\; & \Delta_{\rm g}({\bf n},z_S) - \dfrac{1}{{\cal H}} \partial_r V_\parallel ({\bf n},z_S) \nn
& +\; \int^{\bar{r}_S}_0{d\bar{r} \left(\bar{r} - \bar{r}_S\right)\dfrac{\bar{r}}{\bar{r}_S} \nabla^2_\perp \left(\Phi + \Psi\right) } ,\;
\end{align}
with $\bar{r}_S \,{=}\, \bar{r}(z_S)$ being the background comoving distance at the source redshift $z \,{=}\, z_S$, ${\cal H} \,{=}\, a'/a$ being the comoving Hubble parameter, $a \,{=}\, a(\eta)$ being the cosmic scale factor and, a prime denoting derivative with respect to conformal time $\eta$; the parameters $\Phi$ and $\Psi$ are the well known temporal and spatial Bardeen potentials, the quantity $V_\parallel \equiv {-}{\bf n} \cdot {\bf V} \,{=}\, {-}\bar{n}^i\partial_iV$ is the velocity component along the line of sight, and $V$ is the (gauge-invariant) velocity potential. The operator $\nabla^2_\perp \,{=}\, \nabla^2 - (\bar{n}^i\partial_i)^2 + 2\bar{r}^{-1} \bar{n}^i\partial_i$ is the Laplacian on the plane transverse to the line of sight (the various terms retaining their standard notations). The first term on the right hand side in \eqref{stdDelta_n} is the density term, the second term measures the well known redshift space distortions, and the integral term measures weak lensing.

The relativistic term in the observed overdensity \eqref{Delta_n},
\bea\label{relsDelta_n}
\Delta^{\rm rels}({\bf n},z_S) \;=\; \sum_X \Delta^X_n({\bf n},z_S), 
\eea
is the sum of the individual relativistic corrections $\Delta^X_n$:
\begin{align}\label{Doppler_n}
\Delta^{\rm Doppler}_n \;\equiv\;& \left(b_e - \dfrac{{\cal H}'}{{\cal H}^2} - \dfrac{2}{\bar{r}_S {\cal H}}\right) V_\parallel, \\ \label{ISW_n}
\Delta^{\rm ISW}_n \;\equiv\;& \left(\dfrac{{\cal H}'}{{\cal H}^2} + \dfrac{2}{\bar{r}_S {\cal H}} - b_e\right) \int^{\bar{r}_S}_0{d\bar{r} \left(\Phi' + \Psi' \right) }, \\ \label{timedelay_n}
\Delta^{\rm timdelay}_n \;\equiv\;& \dfrac{2}{\bar{r}_S}\int^{\bar{r}_S}_0{d\bar{r} \left(\Phi + \Psi\right)} , \\ \label{potentials_n}
\Delta^{\rm potentials}_n \;\equiv\;& \left(1 - b_e + \dfrac{{\cal H}'}{{\cal H}^2} + \dfrac{2}{\bar{r}_S {\cal H}}\right) \Phi - 2\Psi \nn
&+\;  \left(3 - b_e\right){\cal H}V + \dfrac{1}{{\cal H}}\Psi', 
\end{align}
where the first term \eqref{Doppler_n} corrects for Doppler effect, the second term \eqref{ISW_n} corrects for the integrated Sachs-Wolfe effect, the second term \eqref{timedelay_n} corrects for time delay, and the last term \eqref{potentials_n} corrects for (local) velocity potential and gravitational potential effects (at the source), respectively. Together, these terms \eqref{Doppler_n}--\eqref{potentials_n} constitute the ``relativistic corrections.'' It should be pointed out that redshift space distortions have both line-of-sight and transverse components; weak lensing, time delay and, Doppler and ISW phenomena, respectively, are all line-of-sight signals; whereas, the density amplitude does not depend on direction---i.e. it only depends on the observation angles, and not on the propagation of the signal.


\section{The Magnification Overdensity}\label{sec:Delta_M}

The cosmic magnification is another important cosmological tool: it allows cosmologists to probe the geometry of the universe on large scales. Henceforth, we use ``magnification overdensity'' to denote the number overdensity of magnified sources, i.e. intrinsically faint sources which require magnification in order to be detected, e.g. as in galaxy size surveys (which measure the angular sizes of the sources; invariably measuring the angular diameter or area distance to the sources). We further take the standard source of cosmic magnification to be weak lensing.

Thus, similarly the observed (relativistic) magnification overdensity \citep[see e.g.][]{Bonvin:2008ni, Jeong:2011as, Duniya:2016ibg, Duniya:2016gcf}, is given by  
\bea\label{Delta_M}
\Delta^{\rm obs}_{\cal M}({\bf n},z_S) \;=\; \Delta^{\rm std}_{\cal M}({\bf n},z_S) + \Delta^{\rm rels}_{\cal M}({\bf n},z_S),
\eea
where we have defined the ``standard'' magnification, as
\bea\label{stdDelta_M}
\Delta^{\rm std}_{\cal M}({\bf n},z_S) \;\equiv\; -{\cal Q}\int^{\bar{r}_S}_0{ d\bar{r}\left(\bar{r} - \bar{r}_S\right) \dfrac{\bar{r}}{\bar{r}_S} \nabla^2_\perp \left(\Phi + \Psi\right)} ,
\eea
with ${\cal Q} \,{=}\, {\cal Q}(z)$ being the cosmic \emph{magnification bias} \citep{Blain:2001yf, Ziour:2008awn, Schmidt:2009rh, Schmidt:2010ex, Jeong:2011as, Liu:2013yna, Camera:2013fva, Duniya:2016ibg, Duniya:2015dpa, Duniya:2016gcf, Hildebrandt:2015kcb}.

Apart from weak lensing as (standard) source of cosmic magnification in an inhomogeneous universe, other sources include Doppler effect, ISW effect, time delay and, (local) potential effects. These effects are corrected for by the relativistic term $\Delta^{\rm rels}_{\cal M}$ which, similar to \eqref{relsDelta_n}, is composed of the components:
\begin{align}\label{Doppler_M}
\Delta^{\rm Doppler}_{\cal M} \;\equiv\;& -2{\cal Q}\left(1 -\dfrac{1}{\bar{r}_S{\cal H}}\right) V_\parallel, \\ \label{ISW_M}
\Delta^{\rm ISW}_{\cal M} \;\equiv\;& 2{\cal Q}\left(1 -\dfrac{1}{\bar{r}_S{\cal H}}\right)\int^{\bar{r}_S}_0{d\bar{r} \left(\Phi' + \Psi' \right)},\\ \label{timedelay_M}
\Delta^{\rm timedelay}_{\cal M} \;\equiv\;& -\dfrac{2{\cal Q}}{\bar{r}_S} \int^{\bar{r}_S}_0{ d\bar{r} \left(\Phi + \Psi\right) }, \\ \label{potentials_M}
\Delta^{\rm potentials}_{\cal M} \;\equiv\;&  2{\cal Q}\Psi + 2{\cal Q}\left(1 -\dfrac{1}{\bar{r}_S{\cal H}}\right) \Phi, 
\end{align}
where \eqref{Doppler_M} gives the Doppler magnification signal, \eqref{ISW_M} gives the ISW magnification signal, \eqref{timedelay_M} gives the (gravitational) time-delay magnification signal, and the term given by \eqref{potentials_M} measures the (gravitational) potential-difference magnification signal, i.e. the magnification owing to the effective gravitational potential well between the source and the observer.


\section{The Dark Sector}\label{sec:Dark}

In this work we take the Universe to be in the late times and dominated by only dark matter (DM) and (fluid) DE, with DE in a non-gravitational interaction with DM: by a reciprocal exchange of energy and momentum. (By virtue of this interaction, DE is referred to as ``interacting'' DE; IDE for short.) We assume that the energy transfer $4$-vectors $Q^\mu_A$ (with $A \,{=}\, m$ for DM and, $A \,{=}\, x$ for DE) are parallel to the DE $4$-velocity, given by \citep{Duniya:2015nva}
\bea\label{trans:Case}
Q^\mu_x \;=\; Q_x u^\mu_x \;=\; -Q^\mu_m,
\eea 
where this implies that there is no momentum transfer in the DE rest frame; with $Q_x$ being the DE energy (density) transfer rate and, $u^\mu_x$ being the DE 4-velocity. The momentum (density) transfer rates, are given by
\bea\label{fm:fx}
f_x \;=\; \bar{Q}_x (V_x - V) \;=\; -f_m,
\eea
where $V$ and $V_x$ are the total and the DE velocity potentials, respectively; the 4-velocities are given by
\bea\label{overDens:Vels}
u^\mu_A &=& a^{-1}\left(1 -\Phi,\, \partial^i V_A\right),\nonumber\\ 
V &=& \dfrac{1}{1+w}\sum_A{\Omega_A\left(1+w_A\right)V_A},
\eea
with $w \,{=}\, \sum_A{\Omega_A w_A}$ being the total equation of state parameter, $\Omega_A \equiv \bar{\rho}_A / \bar{\rho}$ being the density parameter, $\bar{\rho}_A$ being the background energy densities and, $\bar{\rho}$ being the total background energy density. 

We specify the form of IDE, given by \citep{Duniya:2015nva, Duniya:2016gcf, Duniya:2022miz}
\bea\label{Mod2:Q}
Q_x \;=\; \dfrac{1}{3}\xi \rho_x \nabla_\mu u^\mu,
\eea
with $\xi$ (a constant) being the interaction strength; $\rho_x = \bar{\rho}_x + \delta\rho_x$ where $\delta\rho_x$ is the DE (energy) density perturbation, and $\nabla_\mu u^\mu$ measures the total expansion rate (for both background and perturbations). We have taken care to avoid using an energy transfer rate given by $Q_x \,{\propto}\, {\cal H} \rho_x/a$ \citep[see e.g.][]{Ferreira:2014jhn, DiValentino:2019ffd, Abdalla:2022yfr, Zhai:2023yny}, which is a limited approximation of \eqref{Mod2:Q}---as it replaces the total expansion rate with its background term, ${\cal H}$, being the Hubble parameter. (This should lead to different behaviour of the perturbations.)

Equations \eqref{trans:Case}, \eqref{overDens:Vels}, and \eqref{Mod2:Q} then lead to 
\bea\nonumber
Q_x =\; \bar{Q}_x \Big[1 +\delta_x -\Phi -\dfrac{1}{3{\cal H} }\left(3\Phi' - \nabla^2 V\right)\Big] = -Q_m,
\eea
where $\bar{Q}_x \,{=}\, \xi{\cal H}\bar{\rho}_x/a \,{=}\, {-}\bar{Q}_m$ are the DE and the DM background energy density transfer rates, respectively, and $\delta_x \equiv \delta{\rho}_x/\bar{\rho}_x$ is the DE density contrast. (Note that models with $Q_x \,{\propto}\, {\cal H} \rho_x/a$ can only account for up to $\delta_x$ in perturbations.) Our model \eqref{Mod2:Q} \citep[first proposed by][]{Duniya:2015nva}, serves to generalize this class of IDE models. Moreover, we set evolutions such that DM transfers energy to DE:
\bea\label{etd}
\mbox{DM $\to$ DE for}~\xi > 0. 
\eea
\citep[See e.g.][for the full IDE evolution equations.]{Duniya:2015nva}

Furthermore, we adopt the Chevallier-Polarski-Linder parametrization of the DE equation of state parameter $w_x$, given by \citep{Chevallier:2000qy, Linder:2002et} 
\bea\label{wxCPL}
w_x(a) \;=\; w_0 +w_a(1-a),
\eea
where $w_0$ is the value of $w_x$ at today ($a \,{=}\, 1$), and $w_a$ is the slope of $w_x$. For all numerical purposes, we set $w_0 \,{=}\, {-}0.8$ and $w_a \,{=}\, {-}0.2$.


\section{The Large Scale Cosmology}\label{sec:Probes}

\begin{figure*}\centering
\includegraphics[width=15cm,height=10cm]{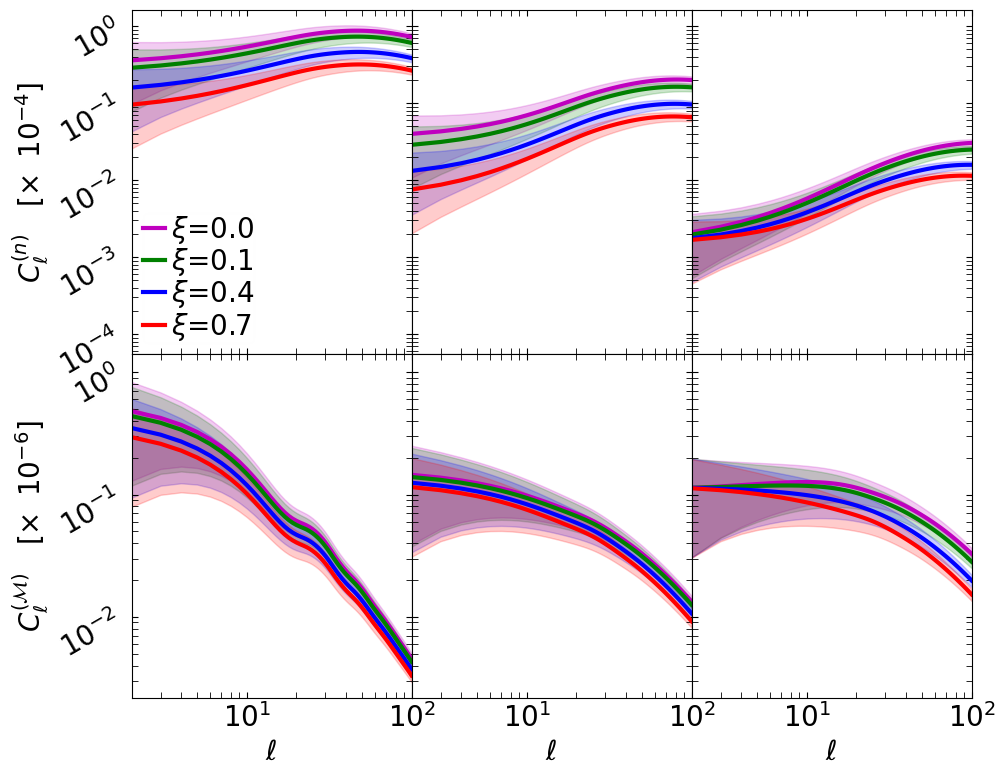}
\caption{The plots of the observed number-count angular power spectrum $C^{(n)}_\ell$ (\emph{top panels}) and the observed magnification angular power spectrum $C^{({\cal M})}_\ell$ (\emph{bottom panels}), as functions of multipole $\ell$: for various values of IDE interaction strength $\xi \,{=}\, 0,\, 0.1,\, 0.4,\, 0.7$. The plots are for the values of source redshifts $z_S \,{=}\, 0.5$ (\emph{left}), $z_S \,{=}\, 1$ (\emph{middle}) and $z_S \,{=}\, 3$ (\emph{right}). The shaded regions show the extent of cosmic variance (assuming a surveyed sky fraction $f_{\rm sky} \,{=}\, 0.75$).}\label{fig:totClsCountsMag}
\end{figure*}

\subsection{The Angular Power Spectra}\label{subsec:Cls}
For the rest of this work, we adopt the spacetime metric with vanishing anisotropic stress, given by
\bea\label{metric} 
ds^2 \;=\; a^2(\eta)\left[-(1+2\Phi)d{\eta}^2 + (1-2\Phi)d{\bf x}^2\right],
\eea
where the vanishing of anisotropic stress implies $\Psi \,{=}\, \Phi$ (see discussions in Secs.~\ref{sec:Delta_n} and~\ref{sec:Delta_M}), and ${\bf x}$ denotes spatial coordinates; with other notations as given in Sec.~\ref{sec:Delta_n}.

The angular power spectrum observed at a source redshift $z_S$, is given by
\beq\label{Cls} 
C^{(I)}_\ell(z_S) = \left(\dfrac{18}{10\pi}\right)^2 \int{dk\, k^2 T(k)^2 P_{\Phi_p}(k) \Big|f^{(I)}_\ell(k,z_S) \Big|^2 },
\eeq
where $T(k)$ is the linear transfer function, and $P_{\Phi_p}$ is the power spectrum of primordial gravitational potential $\Phi_p$. The superscript, $I$, denotes both galaxy redshift number counts ($I \,{=}\, n$) and cosmic magnification ($I \,{=}\, {\cal M}$); with $f^{(n)}_\ell$ and $f^{({\cal M})}_\ell$ being given by \eqref{fn_ell} and \eqref{fM_ell}, respectively:
\begin{align}\label{fn_ell}
f^{(n)}_\ell(k,z_S) =\;& b(z_S) \check{\Delta}_m(k,z_S) j_\ell(kr_S) + j_\ell(kr_S)\dfrac{1}{{\cal H}}\check{\Phi}'(k,z_S) \nn
+\;& \left(b_e - \dfrac{{\cal H}'}{{\cal H}^2} - \dfrac{2}{r_S {\cal H}}\right) \check{V}^\parallel_m(k,z_S) \partial_{kr} j_\ell(kr_S) \nn
-\;& \partial^2_{kr} j_\ell(kr_S)\dfrac{1}{{\cal H}} \partial_r \check{V}^\parallel_m(k,z_S) \nn
+\;& \dfrac{4}{r_S}\int^{r_S}_0{dr\, j_\ell(kr)\, \check{\Phi}(k,r)} \nn
+\;& \left(3 - b_e\right){\cal H}\check{V}_m(k,z_S) j_\ell(kr_S) \nn
+\;& \dfrac{2}{r_S}\int^{r_S}_0{dr \dfrac{\left(r_S - r\right)}{r} \ell\left(1+\ell\right) j_\ell(kr)\, \check{\Phi}(k,r) } \nn
+\;& 2 \left(b_e - \dfrac{{\cal H}'}{{\cal H}^2} - \dfrac{2}{r_S {\cal H}}\right) \int^{r_S}_0{dr\, j_\ell(kr) \check{\Phi}'(k,r) } \nn
-\;& \left(b_e + 1 - \dfrac{{\cal H}'}{{\cal H}^2}  - \dfrac{2}{r_S {\cal H}}\right) \check{\Phi}(k,z_S) j_\ell(kr_S) ,
\end{align}
and,
\begin{align}\label{fM_ell}
f^{({\cal M})}_\ell(k,z_S) =\;& 2{\cal Q}(z_S) \left\lbrace - \dfrac{4}{\bar{r}_S} \int^{\bar{r}_S}_0{d\bar{r}\, j_\ell(k\bar{r}) \check{\Phi}(k,\bar{r})} \right. \nn
&+\; \dfrac{1}{\bar{r}_S} \int^{\bar{r}_S}_0{d\bar{r}\, j_\ell(k\bar{r}) \dfrac{(\bar{r}-\bar{r}_S)}{\bar{r}} \ell(\ell+1)  \check{\Phi}(k,\bar{r})} \nn
&+\; \left(1 - \dfrac{1}{\bar{r}_S {\cal H}}\right) \check{V}^\parallel_m(k,z_S) \partial_{kr}j_\ell(k\bar{r}_S) \nn
&-\; \left(1 - \dfrac{1}{\bar{r}_S {\cal H}}\right) 2\int^{\bar{r}_S}_0{d\bar{r}\, j_\ell(k\bar{r}) \check{\Phi}'(k,\bar{r})} \nn
&+\; \left. \left(2 - \dfrac{1}{\bar{r}_S {\cal H}}\right) j_\ell(k\bar{r}_S) \check{\Phi}(k,z_S) \right\rbrace ,
\end{align}
where $b$ and $\Delta_m$ are the galaxy bias \citep{Desjacques:2016bnm, Baldauf:2011bh, Bartolo:2010ec, Jeong:2011as, Lopez-Honorez:2011emg, Duniya:2016ibg} \citep[see][for an extensive review on galaxy bias]{Desjacques:2016bnm} and the DM comoving overdensity, respectively; $j_\ell$ is the spherical Bessel function, $\partial_{kr} \equiv \partial / \partial(kr)$, and a check denotes division by the value of the gravitational potential $\Phi(k,z_d)$ at the decoupling epoch $z \,{=}\, z_d$. Moreover, we take that on very large scales, where perturbation modes are linear and completely homogeneous and isotropic, galaxies trace the same trajectory as the underlying DM: thus, we set $V^\parallel_{\rm g} \,{=}\, V_\parallel \,{=}\, V^\parallel_m$ (being the line-of-sight peculiar velocity component), and take $b$ as being purely time-dependent.
 
The cosmic variance $\sigma^2_\ell$ at a given redshift, on the angular power spectrum \eqref{Cls}, is given by \citep[see e.g.][]{Maartens:2012rh, Alonso:2015uua, Duniya:2019mpr}
\bea\label{cosmicVar}
\sigma_\ell(z) = \sqrt{\dfrac{2}{\left(2\ell+1\right)f_{\rm sky}}}\,  C^{(I)}_\ell(z),
\eea 
where $f_{\rm sky}$ is the sky fraction covered by a survey. As it is already well known, cosmic variance generally becomes enhanced on very large scales (such as those considered in this work). Physically, this is understandable since the number of galaxy pairs on the given scales becomes limited.

To analyse the the angular power spectra \eqref{Cls}, we initialize the relevant cosmic evolutions at the decoupling epoch, with $1+z_d \,{=}\, 10^3 \,{=}\, a^{-1}_d$, and we use (Gaussian) adiabatic initial conditions for the perturbations \cite[see e.g.][]{Duniya:2013eta, Duniya:2015nva, Duniya:2015dpa, Duniya:2019mpr}. Moreover, for the purpose of our analysis, we use the following parameters for all numerical computations: DE physical sound speed $c_{sx}(z) \,{=}\, 1$, galaxy bias $b(z) \,{=}\, 1$, and magnification bias ${\cal Q}(z) \,{=}\, 1$. We adopt the DM density parameter $\Omega_{m0} \,{=}\, 0.24$, Hubble constant $H_0 \,{=}\, 72~{\rm km\cdot s^{-1}\cdot Mpc^{-1}}$, and speed of light $c \,{=}\, 1$.

In Fig.~\ref{fig:totClsCountsMag} we show the plots of the observed number-count angular power spectrum $C^{(n)}_\ell$ (top panels) and the observed cosmic magnification angular power spectrum $C^{({\cal M})}_\ell$ (bottom panels), with their associated cosmic variance extents (shaded regions): for the values of the dark sector interaction strength $\xi \,{=}\, 0,\, 0.1,\, 0.4,\, 0.7$ (for illustration); at the source redshifts $z_S \,{=}\, 0.5,\, 1,\, 3$, accordingly. We see that in general, while the observed angular power decreases on all scales for galaxy redshift number counts with increasing redshift, the observed angular power for cosmic magnification increases on smaller scales ($\ell \,{\gtrsim}\, 20$), as redshift increases. This is more obvious when the angular power spectrum is, as is customary, multiplied by $\ell(\ell+1)/(2\pi)$: see Fig.~\ref{fig:totClsCountsMag2} in the appendix (included for completeness). This implies that the redshift clustering of sources diminishes with increasing redshift; thus, we will observe fewer sources when we look backwards to earlier epochs. In other words, more and more sources will appear to continue to form on the largest scales as time increases. On the other hand, magnification of cosmic sources will appear to occur more frequently and by large amounts in different regions in the large scale structure as redshift increases. In other words, distant sources will be more magnified than closer sources.

\begin{figure*}\centering
\includegraphics[width=15cm,height=10cm]{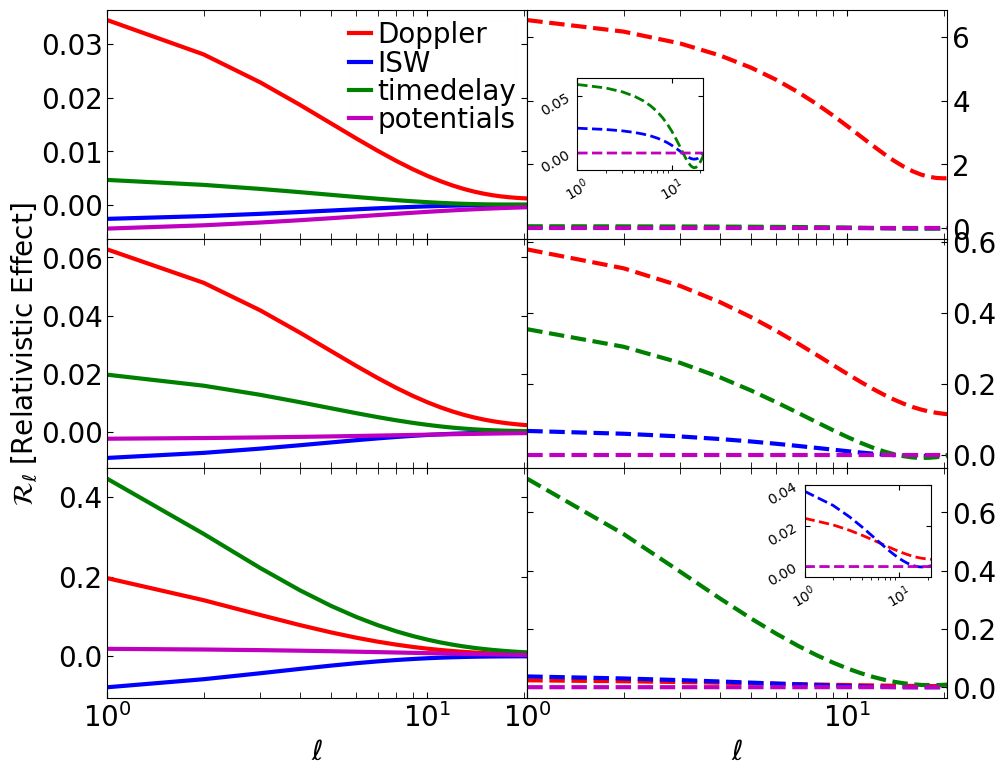}
\caption{The plots of the relativistic effects in the number-count angular power spectrum $C^{(n)}_\ell$ (solid lines) and in the cosmic magnification angular power spectrum $C^{({\cal M})}_\ell$ (dashed lines), as functions of multipole $\ell$: for zero dark sector interaction ($\xi \,{=}\, 0$), at the source redshifts $z_S \,{=}\, 0.5$ (\emph{top panels}), $z_S \,{=}\, 1$ (\emph{middle panels}) and $z_S \,{=}\, 3$ (\emph{bottom panels}).}\label{fig:RelsEffects}
\end{figure*}

Moreover, we see that for both galaxy redshift number counts and cosmic magnifcation, the amplitude of the angular power spectrum is consistently supressed as the value of the dark sector interaction strength increases. This is understandable given our choice of the direction of energy transfer, which is from DM to DE. Thus, this leads to the diminishing of both the clustering of sources and their magnification on the largest scales, at all epochs.

Furthermore, we see the extent of the cosmic variance (shaded regions) for the respective chosen values of IDE strengths. We observe that cosmic variance becomes significant on the largest scales ($\ell \,{\lesssim}\, 10$). These results are consistent with what is already known in the literature: the useful signal will be, in principle, overshadowed by the cosmic variance; thus, will ordinarily not be observed on the largest scales, at the given source redshifts. However, the multi-tracer method \citep[see e.g.][]{Fonseca:2015laa, Alonso:2015sfa, Witzemann:2018cdx} can be used to subdue cosmic variance with large-volume, high-precision future surveys; consequently, allowing the detection of the imprint of IDE (and relativistic effects) in the large scale analysis. Moreover, we observe (in Fig.~\ref{fig:totClsCountsMag}) that for both galaxy redshift number counts and cosmic magnification, the various lines of the angular power spectrum tend to converge on the largest scales as source redshift increases ($z_S \,{\gtrsim}\, 3$). This can be understandable, since DE only becomes significant at late times, i.e. low redshifts ($z \,{<}\, 1$); whereas, at high redshhifts ($z \,{\gtrsim}\, 1$) DM dominates (until radiation epoch) and all pertubation evolutions are similar and linear on large scales.


\subsection{Relativistic Effects}

\begin{figure*}\centering
\includegraphics[width=15cm,height=10cm]{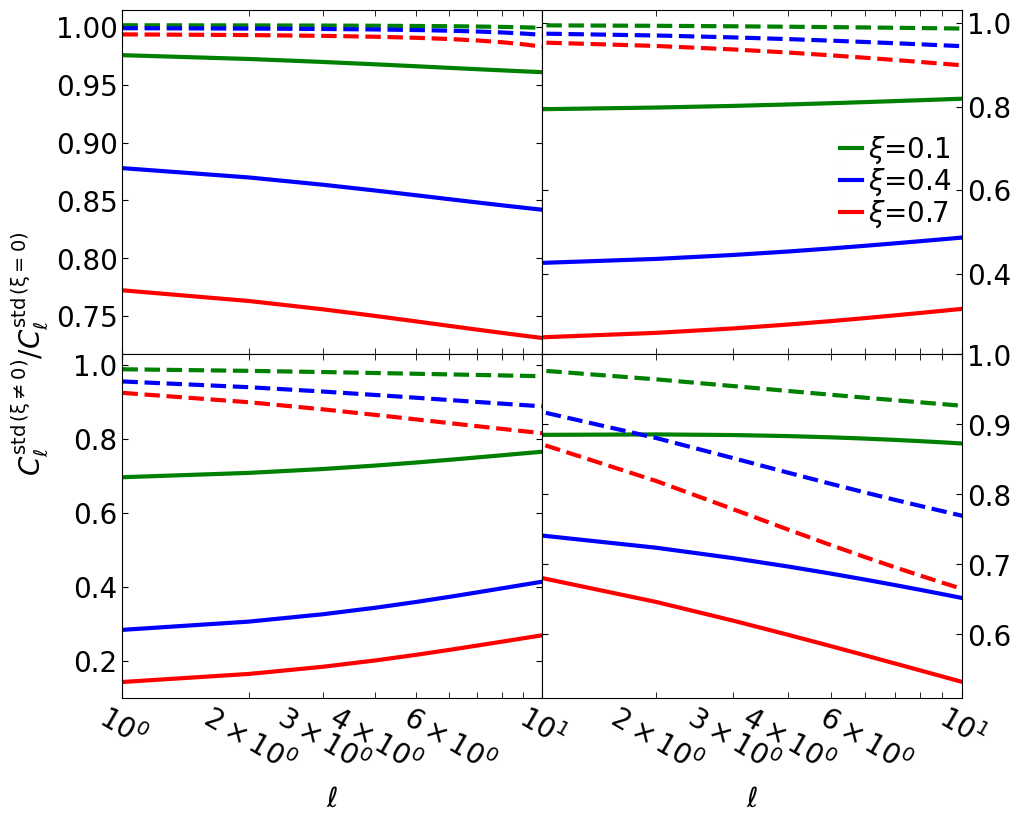}
\caption{The plots of the ratios of the \emph{standard} angular power spectra for IDE ($\xi \,{\neq}\, 0$) to the \emph{standard} angular power spectra for non-IDE ($\xi \,{=}\, 0$): for galaxy redshift number counts (solid lines) and for cosmic magnification (dashed lines), at fixed source redshifts $z_S \,{=}\, 0.1$ (\emph{top left}), $z_S \,{=}\, 0.5$ (\emph{top right}), $z_S \,{=}\, 1$ (\emph{bottom left}), and $z_S \,{=}\, 3$ (\emph{bottom right}); where we used $\xi \,{=}\, 0.1,\, 0.4,\, 0.7$.}\label{fig:stdClsfracs}
\end{figure*}

Here we investigate the effects of the relativistic corrections \eqref{Doppler_n}--\eqref{potentials_n} and \eqref{Doppler_M}--\eqref{potentials_M} in the number-count and the cosmic magnification angular power spectra, repectively. For a given relativistic correction, we determine its total effect, ${\cal R}_\ell$, in the associated angular power spectrum, by
\bea\label{relsEffect} 
{\cal R}_\ell \;\equiv\; \dfrac{C^{(I)}_\ell - C^{(I)({\rm no}\, X)}_\ell}{C^{(I)({\rm no}\, X)}_\ell},
\eea
where $X$ = Doppler, ISW, timedelay, and potentials; with $C^{(I)}_\ell$ being the total angular power spectra with all terms \eqref{Cls}--\eqref{fM_ell} included and, $C^{(I)({\rm no}\, X)}_\ell$ being the angular power spectra with the Doppler, ISW, time-delay, or potentials term excluded, accordingly. These terms are excluded only one at a time, from the associated observed overdensity, and hence from the given $f^{(I)}_\ell$.

In Fig.~\ref{fig:RelsEffects} we show the plots of the relativistic effects \eqref{relsEffect} in the number-count angular power spectrum $C^{(n)}_\ell$ (left panels) and in the cosmic magnification angular power spectrum $C^{({\cal M})}_\ell$ (right panels), for a non-interacting dark sector: at $z_S \,{=}\, 0.5,\, 1,\, 3$, accordingly. We see that in the number-count angular power spectrum, Doppler effect gives the dominant signal at low source redshifts ($z_S \,{\leq}\, 1$). Although Doppler effect gives a significant signal at all redshifts, this signal becomes subdominant at high source redshifts ($z_S \,{>}\, 1$), at which point time delay becomes the dominant signal. The ISW and the potentials effects, respectively, give the least signals: the amplitude of the signal from the potentials effect is lower than that of the ISW effect at $z_S \,{\leq}\, 0.5$. However, the ISW signal gradually surpasses that of the potentials at $z_S \,{>}\, 0.5$---with both signals remaining subdominant to the rest of the signals; even so, giving negative amplitudes. Similalrly, the signal from Doppler effect is the dominant signal from relativistic effects in the cosmic magnification angular power spectrum, at $z_S \,{\lesssim}\, 1$. However, unlike in the number-count angular power spectrum where the amplitude of the Doppler signal is of the same order of magnitude with those of the other effects at all the given values of $z_S$, here the amplitude of the Doppler signal is ${\sim}10^2$ order of magnitude that of the other effects, at $z_S\, {\leq}\, 0.5$. Moreover, unlike in the number-count angular power spectrum, the Doppler signal in cosmic magnification is only significant at $z_S \,{\lesssim}\, 1$, after which point it quickly diminishes drastically to insignificant amplitudes---with respect to the time-delay signal. However, just like in the number-count angular power spectrum, the time-delay signal in the cosmic magnification angular power spectrum becomes dominant at $z_S \,{>}\, 1$; with the rest of the signals becoming insignificant. Although the ISW and the potentials effects, respectively, also give the least signals in the cosmic magnification angular power spectrum, the amplitude of the ISW signal gradually (slightly) surpasses that of the potentials at $z_S \,{>}\, 0.5$: this is the converse of the case in the number-count angular power spectrum. 

In general, the amplitude of the signals from relativistic effects in the number-count angular power spectrum appear to be gradually increasing as $z_S$ increases, whereas the amplitudes in the cosmic magnification angular power spectrum appear to be decreasing with increasing $z_S$. Moreover, in particular, the amplitude of the Doppler and the time-delay signals, respectively, in the cosmic magnification angular power spectrum are much higher than those in the number-count angular power spectrum. The relatively low amplitude of the signals from relativistic effects in the number-count angular power spectrum could be an indication of possible difficulties in measuring these effects with galaxy redshift surveys. Thus, the relatively higher amplitudes in the cosmic magnification angular power spectrum may provide another (relatively better) avenue for measuring the relativistic effects. (However, deeper analysis will be needed to assertain this, which is beyond the scope of this work.) It is already well known that relativistic effects in the angular power spectrum on ultra-large scales are insignificant \citep[see e.g.][]{Maartens:2012rh, Alonso:2015uua, Duniya:2019mpr}, and are below cosmic variance \citep{Alonso:2015uua}. However, a combination of two or more tracers of the DM distribution in the same volume, will suppress cosmic variance and the relativistic effects are able to be detected: the multi-tracer method. 

\begin{figure*}\centering
\includegraphics[width=15cm,height=10cm]{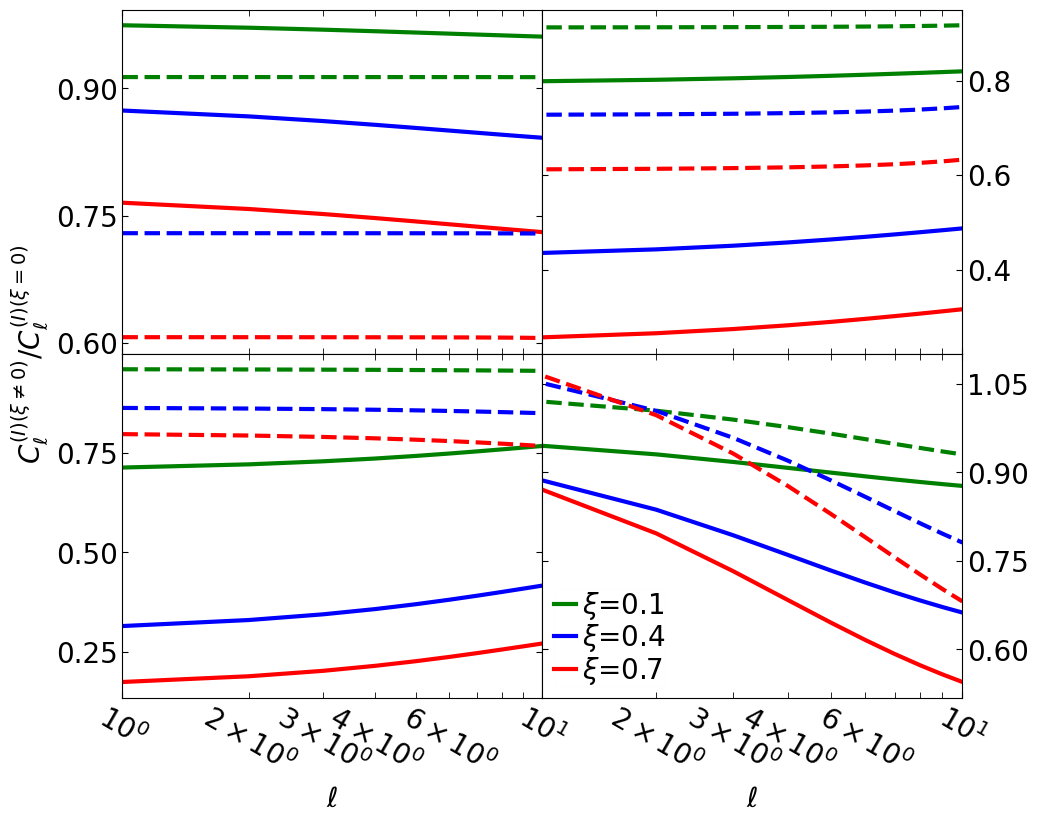}
\caption{The plots of the ratios of the toal angular power spectra for IDE ($\xi \,{\neq}\, 0$) to the total angular power spectra for non-IDE ($\xi \,{=}\, 0$): for galaxy redshift number counts $C^{(n)}_\ell$ (solid lines) and for cosmic magnification $C^{({\cal M})}_\ell$ (dashed lines), at fixed source redshifts $z_S \,{=}\, 0.1$ (\emph{top left}), $z_S \,{=}\, 0.5$ (\emph{top right}), $z_S \,{=}\, 1$ (\emph{bottom left}), and $z_S \,{=}\, 3$ (\emph{bottom right}); with the interaction-strength values $\xi \,{=}\, 0.1,\, 0.4,\, 0.7$.}\label{fig:totClsfracs}
\end{figure*}


\subsection{Dark Sector Interaction}
We further probe the imprint of dark sector interaction in the number-count and the cosmic magnification angular power spectra, respectively.


In Fig.~\ref{fig:stdClsfracs} we show the plots of the ratios of the ``standard'' angular power spectra $C^{{\rm std}}_\ell$, coresponding to the overdensities \eqref{stdDelta_n} and \eqref{stdDelta_M}, accordingly: the ratios of each angular power spectrum with dark sector interaction ($\xi \,{\neq}\, 0$) to that without dark sector interaction ($\xi \,{=}\, 0$). We give these ratios for the values of the dark sector interaction strength $\xi \,{=}\, 0.1,\, 0.4,\, 0.7$, at $z_S \,{=}\, 0.1,\, 0.5,\, 1,\, 3$. We see that the ratios for the cosmic magnification angular power spectrum (dashed lines) have higher amplitudes than those of the number-count angular power spectrum (solid lines), at all the given values of $z_S$. However, the spread between the ratios for the number-count angular power spectrum are larger than those for the cosmic magnification angular power spectrum. In fact, the spread of the ratios for the cosmic magnification angular power spectrum appear to be negligible relative to those of the number-count angular power spectrum. This indicates that the number-count angular power spectrum is more sensitive to changes in the behaviour of IDE or dark sector interactions; thus, suggesting that the ``standard" approximation of the angular power spectrum for galaxy redshift number counts (consisting of density, redshift space distortions, and lensing) will be relatively better in probing IDE models than that of cosmic magnification (consisting of only weak lensing), at all the given values of $z_S$.

In Fig.~\ref{fig:totClsfracs} we show the plots of the ratios of the total angular power spectra $C^{(I)}_\ell$, with dark sector interaction ($\xi \,{\neq}\, 0$) to those without dark sector interaction ($\xi \,{=}\, 0$). (We note that $C^{(I)}_\ell$ contain both the standard terms and the relativistic corrections.) Thus, similar to the case in Fig.~\ref{fig:stdClsfracs}, we give these ratios for the values of the dark sector interaction strength $\xi \,{=}\, 0.1,\, 0.4,\, 0.7$, at $z_S \,{=}\, 0.1,\, 0.5,\, 1,\, 3$. We see that, unlike for the case of the standard andgular power spectrum $C^{\rm std}_\ell$ (Fig.~\ref{fig:stdClsfracs}), the spread of the ratios of the cosmic magnification angular power spectrum (dashed lines) at very low redshifts ($z_S \,{\simeq}\, 0.1$) are larger than those of the number-count angular power spectrum (solid lines), at the same $z_S$. This could be as a result of the signal of the Doppler effect, whose amplitude in the cosmic magnification angular power spectrum is much larger than that in the number-count angular power spectrum, at $z_S \,{\leq}\, 0.5$ (see Fig.~\ref{fig:RelsEffects}). Thus, at these source redshifts ($z_S \,{\lesssim}\, 0.1$) the signals of the relativistic corrections (particularly, the Doppler effect) enables the cosmic magnification angular power spectrum be more sensitive to changes in the parameters of the IDE; consequently, becoming a relatively better probe of the imprint of IDE. However, we see that at source redshifts from $z_S \,{\simeq}\, 0.5$ to $z_S \,{\simeq}\, 1$, the spread between ratios of the number-count angular power spectrum is much larger than that between ratios of the cosmic magnification angular power spectrum. Thus, this suggests that at $0.5 \,{\leq}\, z_S \,{<}\, 3$, the number-count angular power spectrum will be relatively better for probing the imprint of dark sector interaction. At $z_S \,{\geq}\, 3$, the spread between ratios in the number-count and the cosmic magnification angular power spectra, respectively, start to become of the same order of magnitude. The advantage of the number-count angular power spectrum is now lost at these redshifts; hence either the number-count angular power spectrum or the cosmic magnification angular power spectrum will be sufficient to probe IDE models.


\section{Conclusion}\label{sec:Concl}
We performed a qualitative comparative analysis of galaxy redshift number counts and cosmic magnification in a universe with an interacting dark sector, using the angular power spectrum. We investigated the effects of the relativistic corrections to the number-count and the cosmic magnification angular power spectra, respectively, for zero dark sector interaction. We also probed the imprint of dark sector interaction in the number-count and the cosmic magnification angular power spectra, respectively.

We found that the redshift clustering power of galaxy number counts diminishes with increasing redshift. Thus, we will observe fewer sources when we look back into earlier epochs. That is, more and more sources will appear to continue to form on the largest scales as time progresses. On the other hand, the magnification of cosmic sources will appear to occur more frequently and by large amounts in different regions in the large scale structure as we look into the distant past: distant sources will be more magnified than closer sources.

Moreover, the Doppler effect gave the dominant signal in the number-count angular power spectrum, at low redshifts ($z \,{\lesssim}\, 1$). Although the Doppler signal remained significant at all redshifts, it became subdominant at high redshifts ($z \,{>}\, 1$), from where the time-delay signal became the dominant signal. The ISW and the potentials effects, respectively, gave the least signals: the amplitude of the signal from potentials effect was lower than that of the ISW effect at redshifts $z \,{\leq}\, 0.5$. However, the ISW signal gradually surpasses that of the potentials effect, at redshifts $z \,{>}\, 0.5$.

Similalrly, the Doppler effect gave the dominant relativistic signal in the cosmic magnification angular power spectrum, at low redshifts ($z \,{\lesssim}\, 1$). However, unlike in the number-count angular power spectrum, the amplitude of the Doppler signal in cosmic magnification was found to be much enhanced relative that of the other signals, at redshifts $ z\, {\leq}\, 0.5$. Also, the Doppler signal in cosmic magnification is only significant at low redshifts ($z \,{\lesssim}\, 1$). However, similar to the case in the number-count angular power spectrum, the time-delay signal in the cosmic magnification angular power spectrum becomes dominant at high redshifts ($z \,{>}\, 1$). Although the ISW and the potentials effects, respectively, gave the least signals in the cosmic magnification angular power spectrum, the amplitude of the ISW signal gradually surpasses that of the potentials effect at redshifts ($z \,{>}\, 0.5$): this is the converse of the case in the number-count angular power spectrum. 

Moreover, our results suggest that measuring relativistic effects in the cosmic magnification angular power spectrum will be relatively better than in the number-count angular power spectrum. Nevertheless, more analysis will be needed to assertain this. It has already been pointed out \citep[see e.g.][]{Duniya:2019mpr} that a multi-tracer analysis will be needed to detect the relativistic effects in galaxy redshift number counts. Thus, in the light of multi-tracer analysis, cosmic magnification offers a new (and relatively better) avenue to measure relativistic effects in the large scale structure. 

Furthermore, we found that without relativistic effects, the number-count angular power spectrum holds the potential to be relatively better in probing the imprint of IDE than the cosmic magnification angular power spectrum, at all redshifts. On the other hand, relativistic corrections in the observed overdensity will enable the cosmic magnification angular power spectrum to be a relatively better probe of the imprint of IDE, at low redshifts ($z \,{\lesssim}\, 0.1$). At redshifts, $0.5 \,{\lesssim}\, z \,{<}\, 3$, the number-count angular power spectrum will be relatively better at probing the imprint of dark sector interactions. At high redshifts ($z \,{\geq}\, 3$), our results suggest that either the number-count angular power spectrum or the cosmic magnification angular power spectrum will be sufficient in probing the imprint of IDE.

Lastly, it should be emphasized that the analysis presented in this work is mainly qualitative; hence the results mainly illustrate the potential of galaxy redshift number counts and cosmic magnification as cosmological probes. In order to properly answer the question in the title of this paper, a rigorous quantitative analysis, which takes cosmic variance (among other factors, see the introduction) into account and employs multi-tracer techniques, will be required. However, as previously stated, we retain the given title for the sake of arousing discussion in the community towards finding a definitive answer.


\section*{Acknowledgements}
We thank the refree for useful comments. We acknowledge the use of high performance computing facilities of the University of Botswana, for the numerical computations in this work. MK has received funding from the PanAfrican Planetary and Space science Network (PAPSSN). PAPSSN is founded by the Intra-Africa Academic Mobility Scheme of the European Union under the grant agreement No. 624224.

\section*{Data Availability}
Data sharing is not applicable to this article, as no observational datasets were generated or analysed in the current study. 




\bibliographystyle{mnras}
\bibliography{counts_vs_magnification} 


\appendix 

\section{The $C_\ell$'\lowercase{s} scaled up}\label{App:fig5}

\begin{figure*}\centering
\includegraphics[width=15cm,height=10cm]{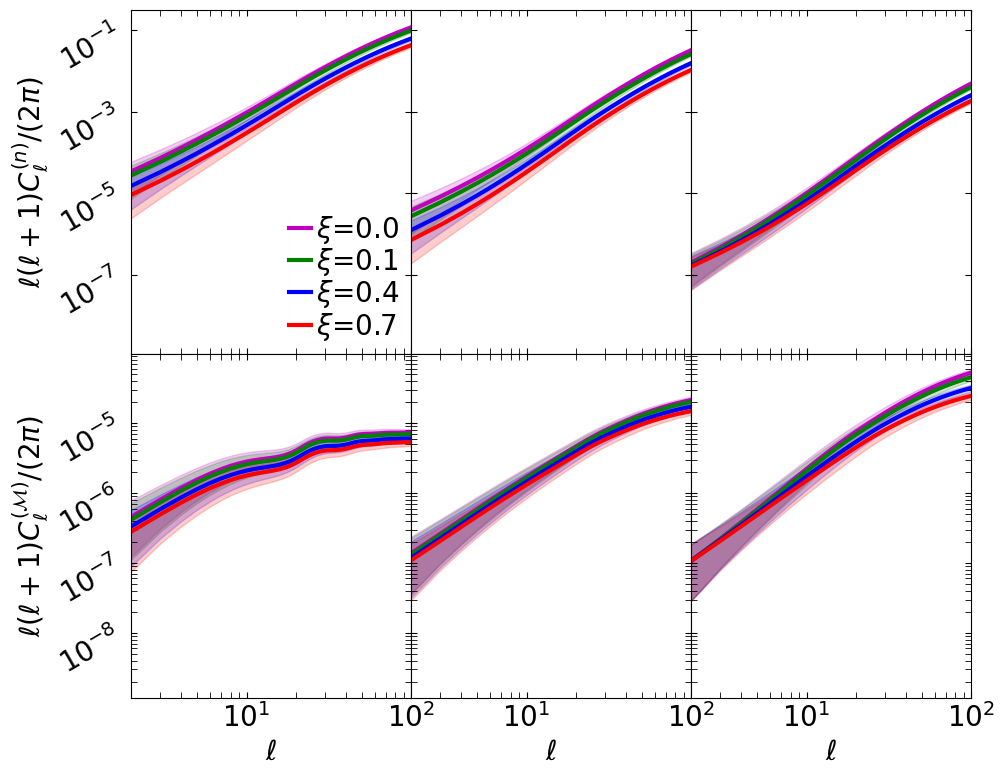}
\caption{The plots of the observed number-count angular power spectrum, as $\ell(\ell+1) C^{(n)}_\ell / (2\pi)$ (\emph{top panels}), and the observed magnification angular power spectrum, as $\ell(\ell+1) C^{({\cal M})}_\ell / (2\pi)$ (\emph{bottom panels}), as functions of multipole $\ell$: for various values of IDE interaction strength $\xi \,{=}\, 0,\, 0.1,\, 0.4,\, 0.7$. The plots are for the values of source redshifts $z_S \,{=}\, 0.5$ (\emph{left}), $z_S \,{=}\, 1$ (\emph{middle}) and $z_S \,{=}\, 3$ (\emph{right}). The shaded regions show the extent of cosmic variance. This figure is equivalent to Fig.~\ref{fig:totClsCountsMag}: the only difference is the factor of $\ell(\ell+1) / (2\pi)$.}\label{fig:totClsCountsMag2}
\end{figure*}

For completeness, in Fig.~\ref{fig:totClsCountsMag2} we show the plots of the angular power spectra $C^{(n)}_\ell$ and $C^{({\cal M})}_\ell$ of Fig.~\ref{fig:totClsCountsMag}, but here with each spectrum being scaled by the factor $\ell(\ell+1) / (2\pi)$. The advantage of this is that it reveals the high-multipole behaviour of the power spectra, as source redshift increases.

\bsp	
\label{lastpage}
\end{document}